%% file: moiseenko_main.tex
\documentclass[multphys,vecphys]{svmult}

\usepackage{makeidx}     
\usepackage{graphicx}    
\usepackage{multicol}    

\makeindex             

\begin{document}

\title*{Magnetorotational mechanism of supernova type II exlosion}
\author{Moiseenko S.G. \inst{1}\and
Bisnovatyi-Kogan G.S.\inst{1}\and
Ardeljan N.V.\inst{2}
}
\institute{
Space Research Institute, Profsoyuznaya str. 84/32 Moscow 117997,
Russia \texttt{gkogan@iki.rssi.ru, moiseenko@iki.rssi.ru}
\and
Department of Computational Mathematics and
Cybernetics, Moscow State University, Vorobjevy Gory, Moscow B-234 Russia
\texttt{ardel@cs.msu.su}
}
%
%
\maketitle

\abstract
\label{abs:1}
Results of 2D simulations of the magnetorotational  mechanism
of supernova type II are presented. Amplification of toroidal
magnetic field of the star due to differential rotation of the
star leads to the transformation of the rotational (gravitational)
energy to the energy of the supernova explosion. In our simulation
the energy of the explosion is $1.12 \cdot 10^{51}erg$. The explosion ejects
about $0.11 M_\odot$.

\section{Magnetorotational mechanism}
\label{sec:1}
The magnotorotational supernova  (MRS) explosion model was suggested in \cite{bk70}.
The idea of MRS consists of getting explosion  energy from the rotational (gravitational)
energy of the collapsed magnetized massive star. 1D numerical simulation of the MRS mechanism
has been made in \cite{bkpopsam},\cite{ardbkpop}. We have made 2D numerical simulation of the
MRS using specially developed implicit conservative Lagrangian scheme on triangular
grid with grid reconstruction. Our results show that MRS leads to the
energy output of the $1.12\cdot 10^{51}$erg and ejection of $0.11M_\odot$.

Core collapse of a star leads to formation of the
rapidly (almost rigidly) rotating neutron core
and differentially rotating vast envelope.
The explosion energy for the supernova is taking form the
rotational (gravitational) energy of the magnetized star. The magnetic
field plays the role of the "transmission belt"
for the rotational (gravitational) energy to the energy of
the supernova explosion. Toroidal component of the magnetic field
is amplifying with time due to the differential rotation of the star.
When force produced by magnetic pressure substantially changes the local balance of forces
a compression MHD wave appears
and goes through the star's envelope outwards. Moving along steeply decreasing density
profile this wave transforms to the MHD shock which produces supernova explosion.

\section{Formulation of the problem}
\label{sec:2}
\subsection{Basic equations}
\label{sec:3}
Consider a set of magnetohydrodynamical equations with
self\-gra\-vi\-ta\-tion and with infinite conductivity:
\begin{eqnarray}
\frac{{\rm d} {\bf x}} {{\rm d} t} = {\bf u}, \quad
\frac{{\rm d} \rho} {{\rm d} t} +
\rho \nabla \cdot {\bf u} = 0,  \nonumber
\end{eqnarray}
\begin{eqnarray}
\rho \frac{{\rm d} {\bf u}}{{\rm d} t} =-\nabla
\left(p+\frac{{\bf H} \cdot {\bf H}}{8\pi}\right) +
\frac {\nabla \cdot ({\bf H} \otimes {\bf H})}{4\pi} -
\rho  \nabla \Phi,
\nonumber
\end{eqnarray}
\begin{eqnarray}
\rho \frac{{\rm d}}{{\rm d} t} \left(\frac{{\bf H}}{\rho}\right)
={\bf H} \cdot \nabla {\bf u},\quad
\Delta \Phi=4 \pi G \rho,
\nonumber
\end{eqnarray}
\begin{eqnarray}
\rho \frac{{\rm d} \varepsilon}{{\rm d} t} +p \nabla \cdot {\bf u}+\rho F(\rho,T)=0,
\nonumber
\end{eqnarray}
where $\frac {\rm d} {{\rm d} t} = \frac {\partial} {
\partial t} + {\bf u} \cdot \nabla$ is the total time
derivative, ${\bf x} = (r,\varphi , z)$, ${\bf u}$ is velocity
vector, $\rho$ is density, $p$ is pressure,  ${\bf
H}=(H_r,\> H_\varphi,\> H_z)$ is magnetic field vector, $\Phi$ is
gravitational potential, $\varepsilon$ is internal energy, $G$ is
gravitational constant,  ${\bf H} \otimes {\bf H}$ is tensor
of rank 2, $F(\rho,T)$ is the rate of neutrino losses, other notations are standard..

Axial symmetry ($\frac \partial {\partial
\varphi}=0$) and symmetry to the equatorial plane ($z=0$) are
assumed.

\subsection{Equations of state}
\label{sec:4}

The equations of state $P(\rho,T)$ \cite{abkpch}. includes approximation of the
tables from  \cite{bps, mjb} for the cold degenerate matter $P_0(\rho)$:
$
  P \equiv P(\rho,T)=P_0(\rho)+\rho \Re T + \frac {\sigma T^4} {3},
$

\begin{equation}\label{pressure}
P_0(\rho)=\left\{
\begin{array}{rclcccccc}
P_0^{(1)}&=&b_1\rho^{1/3}/(1+c_1\rho^{1/3}),& &at&\rho&\leq&\rho_1,&\\
P_0^{(k)}&=&a\cdot 10^{b_k({\textrm{lg}}\rho-8.419)^{c_k}}&at \> \rho_{k-1}&\leq&\rho&\leq&\rho_k,&k=\overline{2,6}\\
\end{array}
\right.
\end{equation}
$
b_1=10.1240483,\>    c_1=10^{-2.257},\>  \rho_1=10^{9.419}, \>
b_2=1.0,  \>         c_2=1.1598,  \>     \rho_2=10^{11.5519},\>
b_3=2.5032, \>       c_3=0.356293,\>     \rho_3=10^{12.26939}, \>
b_4=0.70401515, \>   c_4=2.117802,\>     \rho_4=10^{14.302}, \>
b_5=0.16445926, \>   c_5=1.237985,\>     \rho_5=10^{15.0388},\>
b_6=0.86746415, \>   c_6=1.237985,\>     \rho_6\gg \rho_5, \>
a=10^{26.1673.}
$

Here $\rho$ is a total mass-energy.
The energy of the unit mass is defined as:
$
\varepsilon=\varepsilon_0(\rho)+\frac{3}{2}
\Re T +\frac{\sigma T^4}{\rho},
$
where $\Re$ - gas constant, $\sigma$ - constant radiation density,
and $ \varepsilon_0(\rho)=\int\limits_0^\rho \frac{P_0(\tilde{\rho})}{\tilde{\rho}^2}\textrm{d}\tilde{\rho}
$.

The neutrino losses from Urca processes are defined by relation \cite{iin}:
\begin{eqnarray}\label{urca}
  f(\rho,T)=\frac {1.3 \cdot 10^9 {\textrm {\ae}}(\overline{T})\overline{T}^6}
  {1+(7.1\cdot 10^{-5}\rho \overline{T})^{\frac{2}{5}}}\quad
  {\textrm {erg}} \cdot {\textrm{g}}^{-1} \cdot {\textrm {c}}^{-1},\\
{\textrm {\ae(T)}}=\left\{
\begin{array}{rclcccccc}
1,&\overline{T}<7,\\
664.31+51.024 (\overline{T}-20), & 7\leq \overline{T} \leq 20,\\
664.31, & \overline{T}>20,\> \overline {T}=T\cdot 10^{-9}.
\end{array}
\right.
\end{eqnarray}
Neutrino losses from pair annihilation, photo-production
and plasma were also taken into account. These type of
the neutrino losses have been approximated by the interpolation formulae from
\cite{schindler} :
$
  Q_{tot}=Q_{pair}+Q_{photo}+Q_{plasm}
$
These three terms  can be written in the following general form:
\begin{equation}\label{schindler1}
  Q_d=K(\rho,\alpha)e^{-c\xi}\frac{a_0+a_1\xi+a_2\xi^2}{\xi^3+b_1\alpha+b_2\alpha^2+b_3\alpha^3}.
\end{equation}
For $d=pair$, $K(\rho,\alpha)=g(\alpha)e^{-2\alpha}$,
$
g(\alpha)=1-\frac{13.04}{\alpha^2}+\frac{133.5}{\alpha^4}+\frac{1534}{\alpha^6}+\frac{918.6}{\alpha^8};
$
For $d=photo$, $K(\rho,\alpha)=(\rho/\mu_Z)\alpha^{-5};$
For $d=plasm$, $K(\rho,\alpha)=(\rho/\mu_Z)^3;$
$
\>\xi=\left(\frac{\rho / \mu_Z}{10^9}\right)^{1/3} \alpha.
$
Here $\mu_Z=2$ is number of nucleons per electron. Coefficients $c,\> a_i,\> b_i$
for different $d$ are given in \cite{schindler}.
\begin{figure}
\centering
\includegraphics{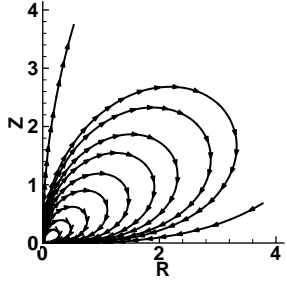}
\caption{Initial poloidal magnetic field.}
\label{mag0}
\end{figure}
The general formula for  neutrino losses in nontransparent star has been written in the following form:
$
  F(\rho,T)=(f(\rho,T)+Q_{tot})e^{-\frac{\tau_\nu}{10}}.
$
The multiplier $e^{-\frac{\tau_\nu}{10}}$,
where $\tau_\nu=S_\nu n l_\nu$
restricts neutrino flux for non zero depth to neutrino interaction with matter $\tau_\nu$.
The cross-section for this interaction $S_\nu$ was represented in the form
$S_\nu=\frac{10^{-44}T^2}{(0.5965 \cdot 10^{10})^2}$, the nucleons concentration is:
$n=\frac{\rho}{m_p}$,
$m_p=1.67 \cdot 10^{-24} {\textrm{g}}$. The characteristic length scale $l_\nu$
which defines the depth for the neutrino absorbtion, was taken
to be equal to the characteristic length of the density variation, as
$
l_\nu=\frac{\rho}{
|\nabla \rho|}=\frac{ \rho}{\left((\partial \rho /\partial r)^2+
(\partial \rho / \partial z)^2\right)^{1/2}}.
$
This value monotonically decreases when moving to the outward boundary,
its maximum is in the center.
It approximately determines to the depth of the
neutrino absorbing matter. The
multiplier $1/10$ was applied because in the
degenerate matter of the hot neutron star
only part of the
nucleons with the energy near Fermi energy
(it was taken   $\approx1/10$) takes part in the neutrino
interaction processes.

\section{Results}
For the numerical simulations we have used implicit Lagrangian difference scheme on triangular
grid with grid reconstruction. For the description of the applied numerical
method see, for example, \cite{arkos} and references therein.
The number of knots of the triangular grid was about 5000.
\begin{figure}
\centerline{\includegraphics{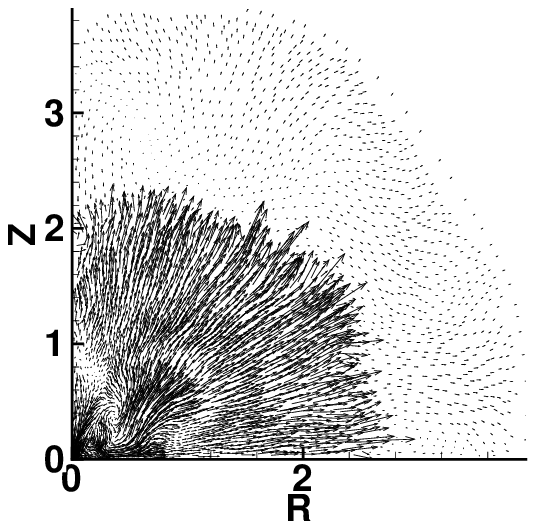} \includegraphics{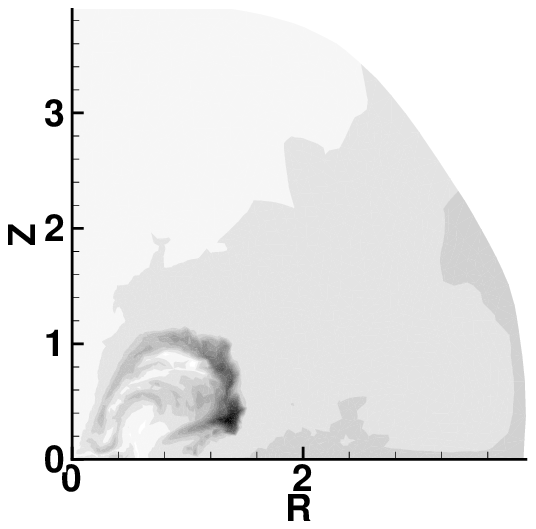}}
\caption{Velocity field (left plot) and specific angular momentum $v_\varphi r$ (right plot)
 at $t=0.191$s
after turning on the magnetic field. The darker parts at
the right plot correspond to the larger angular momentum}
\label{expl}       
\end{figure}
As a first stage of the MRS mechanism we have calculated a collapse of rotating star,
leading to  differentially rotating configuration.
After the collapse the star (core of evolved massive star) consists of a almost rigidly rotating neutron
star with radius $\sim 10$km
which rotates with the period $\sim 0.001$ sec, and vast differentially rotating envelope .
\begin{figure}
\centering
\includegraphics{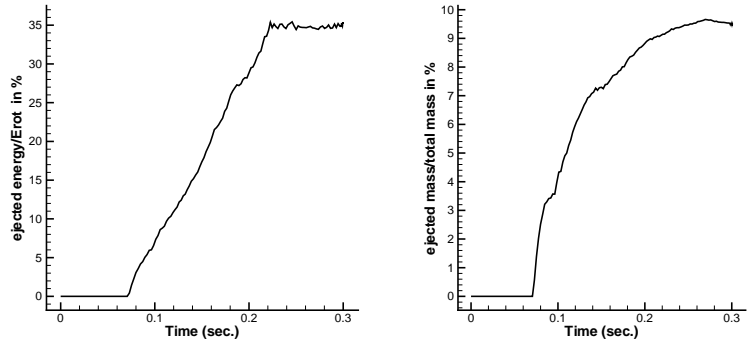}
\caption{Time evolution of the (ejected energy)/(rotational energy) in \%  (left plot)
and (ejected mass)/(total mass) in \%. (right plot).}
\label{eject}       
\end{figure}
After formation of the differentially rotationg configuration the initial poloidal
 magnetic field was switched on (Fig.\ref{mag0}).

The energy of the initial magnetic field was $E_{mag 0}=10^{-6}E_{grav}$.
 Where $E_{mag 0}$ - is the energy of the
initial magnetic field, $E_{grav}$ - is the
gravitational energy of the collapsed star.

Due to the differential rotation the toroidal component of the
magnetic field is increasing  with time. The magnetic pressure
growths and produces compression MHD wave, which moves through
envelope with steeply decreasing density. Soon after appearing
this wave transforms into the MHD shock, which throws away part of
the matter of the envelope. The MHD shock front is clearly seen
at the velocity field plot (Fig.\ref{expl} - left plot).
The magnetic field transmits angular momentum of the neutron star outwards
(Fig.\ref{expl} - right plot).

Results of our simulations show that the energy of the supernova explosion is
about $1.12\cdot 10^{51}erg$ (35\% of the rotational energy of the star).
The explosion ejects about $0.11M_\odot$ ($\sim 9.7 \%$ of the mass of the star).
At the Fig.\ref{eject}
the time evolution of the ejected mass and ejected energy are presented.

Detailed description of the results of the simulations of the MRS will
be published elsewhere.

The authors G.S.B.-K. and S.G.M would like to thank the Organizing Committee for the
support and hospitality.
This work was partially supported by the grant RFBR 02-02-16900.
\input{moiseenko_ref}



\printindex
\end{document}

%% file: moiseenko_ref.tex
%
%

%
%